\documentclass[12pt]{article}
\usepackage{fullpage}
\usepackage{times}
\usepackage{multirow}
\usepackage{comment}
\usepackage{setspace}
\usepackage{graphicx}
\usepackage{natbib}

\setcitestyle{authoryear,round,aysep={}}
\title{Explorations in Statistics Research: \\ 
An Approach to Expose Undergraduates to\\ 
Authentic Data Analysis}

\author{Deborah Nolan\\ Berkeley, CA 94720-3860\footnote{Deborah Nolan is a Professor, Department of Statistics, University of California, 367 Evans Hall MC 3860, Berkeley CA 
94720-3860, (email: \texttt{deborah\_nolan@berkeley.edu})}\\
 Duncan Temple Lang\\Davis, CA 95616\footnote{Duncan Temple Lang is 
 Director of the campus Data Sciences Initiative and a Professor,
 Department of Statistics, University of California, 4210 Mathematical Sciences Building,
 One Shield Ave, Davis, CA 95616 (email: \texttt{dtemplelang@davis.edu}). 
The authors gratefully acknowledge support from the National Science Foundation grant 
DMS-0840001.}}

\begin{document}

\maketitle

\newpage

\begin{abstract}
The Explorations in Statistics Research workshop is a one-week NSF-funded summer program that
introduces undergraduate students to current research problems in applied statistics.  The goal of
the workshop is to expose students to exciting, modern applied statistical
research and practice,  with the ultimate aim of interesting them in 
seeking more training in statistics at the undergraduate and
graduate levels.  The program is explicitly designed to
engage students in the connections between authentic domain problems and the
statistical ideas and approaches 
needed to address these problems, which is an important aspect of statistical thinking that is 
difficult to teach and sometimes lacking in our methodological courses and programs. 
Over the past nine years, we ran the workshop six times and a similar program in
the sciences two times.  We describe the program, summarize feedback from 
participants, and identify the key features to its success.  
We abstract these features and provide a set of recommendations for how
faculty can incorporate important elements into their regular courses.
\end{abstract}

\vspace*{.3in}

\noindent{\textsc{Key Words}}: statistical problem solving, visualization, 
data analysis pipeline, pedagogy, co-curricular activity.

\doublespacing

\section{INTRODUCTION}

The Explorations in Statistics Research (ESR) workshop is a week-long summer program where
undergraduates work closely with statisticians and graduate students to analyze data from important
research problems at the frontier of applied statistics.  The aim of the program is to give students
an understanding of and experience with the role of statistics in scientific discovery with the goal of
encouraging them to pursue advanced studies in statistical science.
The students are guided through the process of using statistics to address interesting scientific
and social questions. In the workshop, they experience how statisticians work and reason about an
authentic problem in a science or industry domain. 

This one-week program attempts to bridge some of the gap between the teaching of statistics and
the modern practice of statistics by exposing students to the 
interplay between a question in a scientific area and the way statistics can address the 
question.  
The workshop gives students exposure to how a statistician frames a research question 
in statistical terms, and they gain first hand experience with how to explore relevant data, 
understand the statistical issues, and use statistical methods to address the scientific 
question.
\cite{bib:Speed} describes the importance of this aspect of our field and why it should
be a central part of statistics education:
\begin{quote}
The interplay between questions, answers and statistics seems to me to be something which should interest teachers of statistics, for if students have a good appreciation of this interplay, 
they will have learned some statistical thinking, not just some statistical methods. 
Furthermore, I believe that a good understanding of this interplay can help resolve many of the 
difficulties commonly encountered in making inferences from data.
\end{quote}
The importance of this interplay in educating our students features prominently   
in the American Statistical Association's
2014 Curriculum Guidelines for Undergraduate Programs in Statistical Science
\citep{bib:ASAGuidelines}.
There, the first guiding principle is the 
``scientific method and its relation to the statistical problem solving cycle,'' 
and the guidelines state (p.6):
\begin{quote}
All too often, undergraduate statistics majors are handed a ``canned" data set and told
to analyze it using the methods currently being studied. ...  Students need practice developing
a unified approach to statistical analysis and integrating multiple methods in an 
iterative manner. ... Students need to see that the discipline of statistics is more
than a collection of unrelated tools (or methods); it is a general approach to 
problem solving using data.
\end{quote}

In this paper, we present the ESR workshop and draw lessons from our experience
with the program, which we hope provide insights and ideas for addressing this 
central aspect of statistics training. The ESR has evolved over the years 
as we experimented with different approaches and gathered student feedback. 
ESR began in 2005, organized by Hansen, Nolan, and Temple Lang, 
and was offered six times between 2005 and 2012.  All together, 21 researchers
worked with a total of 146 undergraduate participants and 
approximately 45 graduate students, teaching faculty, and organizers.
We first describe the core of the program, including a description of how one 
researcher engaged students in her research. 
Next, we review the impact of the ESR as reported by the undergraduate and graduate participants,  
summarize the key elements of success,  
and present ideas for how these features can be adopted in the statistics major.
These proposed changes to what and how we teach will equip our students with 
essential skills in statistical thinking with data.

\section{THE INVESTIGATIVE MODEL}

The Explorations in Statistics Research workshop offers a strong scientific program led by 
research statisticians working at the frontiers of modern applications in science, policy,
industry, political science, etc.    
It exposes undergraduates to research and the associated data analysis process 
that the students often do not experience until a capstone project at the end of their major, 
past the point when they are deciding on a major or whether to apply to graduate school.
The core of the one-week workshop consists of three two-day data analysis projects. 
Each two-day topic is led by a researcher who organizes the activities to engage the students on problems related to his or her current work. 
(Table~\ref{tab:topiclist} lists the topics, researchers, and their institutions for all offerings of ESR.) 
The researcher provides data for analysis and prepares short talks and computer investigations 
where the students are introduced to the material in stages. 
Through hands-on data analysis, students experience statistical research, 
from exploratory data analysis (EDA) to modeling to conclusions.

\begin{table}
\begin{tabular}{|l|l|l|p{1.45in}|}
\hline
\textbf{Year} & \textbf{Topic} & \textbf{Researcher} & \textbf{Affiliation} \\
\hline
2012 & Phylogenetic trees & Katie Pollard & UC San Francisco \\
2012 & Twitter & Mark Hansen & Columbia\\
2011 & Market Segmentation &	Andreas Buja & U Pennsylvania \\
2011	& Data Visualization in Journalism &	Amanda Cox & New York Times\\
2011 & Health Care & David Madigan  & Columbia\\
 &  &  Patrick Ryan &  OMOP\\ 
2010 & Updating Search Engine databases & Carrie Grimes & Google\\
2010	 & Cosmology and computer experiments	& Dave Higdon & LANL\\
2010 & Regional climate model	 & Steve Sain & NCAR\\
2009	 & Voting irregularities in Florida &  Jasjeet Sekhon & UC Berkeley\\
2009 & Combining Global Climate Models	 & Claudia Tebaldi	 & U British Columbia \\
2009	 & Building a recommender system & Chris Volinsky & AT\&T Research\\
2006	 & Pollution and Mortality & Francesca Dominici & Johns Hopkins\\
 & & Roger Peng & Johns Hopkins\\
2006	 & Cosmology and the expanding universe & Chris Genovese & Carnegie Mellon\\
2006	 & Bayesian approaches to Geo-Location & David Madigan & Rutgers\\
2005	 & Wireless Geo-Location	& Diane Lambert & Google \\
2005 & Extreme weather events & Doug Nychka & NCAR\\
2005	 & Traffic patterns and problems  & John Rice & UC Berkeley\\
2005	 & Genetics and clustering	 & Terry Speed & UC Berkeley\\
2005 & Image Analysis & Ying Nian Wu & UC Los Angeles\\
\hline
\end{tabular} 

\caption{Research topics from the Explorations in Statistics Research Workshop, 2005-2012.
Listed here are the topics addressed, the researcher who organized and presented each topic, 
and the researcher's institution. (Note: LANL is the Los Alamos National Laboratory, 
NCAR is the National Center for Atmospheric Research, and
OMOP is the  Observational Medical Outcomes Partnership.)}
\label{tab:topiclist}
\end{table}

Throughout the workshop, students have multiple opportunities to converse with the researcher 
about their insights and ideas, both collectively and individually. 
The activities carefully build upon each other, beginning with simple EDA and advancing 
to the application of modern statistical methods.  
The three-to-one student to ``teacher" ratio enables the students to be creative without 
getting bogged down by computational issues of implementation.  
The workshop emulates a research/work environment involving group work and 
frequent presentation and discussion of ideas.

In general, the topic begins with the researcher providing a high-level description of 
his or her problem and the associated measurements and data, and what the ultimate goal is. 
The length and depth of the introductory material depends somewhat on the immediacy of the underlying data.
The students ask questions and the researcher guides a discussion that allows
the students to think about aspects of the problem and potential issues and directions.
This is followed by a breakout session in which the students explore, and become familiar with, the
data, primarily through visualization.
The entire group comes back together to present some of their findings and discuss new issues
uncovered during their explorations.  
These presentations are very short, in one to two minutes a student or group of students present
a statistical graph they have created and describe their findings.
The researcher guides the discussions and connects students' findings to 
the larger data analysis problem and next steps.  
Given some further questions, the students return to group work and 
continue with their analysis in greater depth.
With this approach, students quickly uncover the main structure of the data, they employ
computationally powerful, yet intuitive, modern statistical methods, and they delve deeper to begin
to unlock the answers to important research questions. 

As the two days progress, we regularly transition between full-group discussions with
the researcher and  small-group work with brain storming and exploring.
The researcher introduces one or more statistical approaches 
to the primary problem(s),  and the students work with the data using these methods.
The students become more adept at asking questions
that are grounded in statistical concepts and answered through computation and visualization. 
In the last afternoon, the students present their final discoveries, again in the form of a plot. 
The researcher moderates the event and provides feedback,
pointing out different possible directions, anomalies, etc. and ties students' findings to
the original question.
Then the researcher concludes by presenting 
a perhaps more sophisticated solution and additional work on this
and other current research problems,
connecting his/her work to the students' findings and more fully demonstrating their approach 
to the problem.

To get a sense of how a topic might actually unfold, we provide a description of 
Carrie Grimes' topic on updating a Web cache, 
i.e., Google's index of all of the pages on the Web.
In the description below, we attempt to give a flavor for the sequence
of activities and to make explicit how the ESR approach is quite different from a
typical classroom learning environment. 

\paragraph{Updating a Web Cache, Carrie Grimes, Google Research.}
In an introductory session on the morning of the first day, 
Grimes described how search engines keep current their database of Web pages.  
The main issue she wanted the students to focus on was determining how often a Web site 
should be revisited by the search engine to ensure that the index is up-to-date.  
Google wants the index to always have the most up-to-date version of the page.  
However, it's not possible to check if each page has changed
every second, minute or even hour.  This is the crux of the problem on which Grimes focused.  
She also introduced several additional issues that arise when trying to index the World Wide Web,
which helped put the research question into a larger context.  
At the end of her introduction, Grimes described the data that had been collected to study this
problem: thousands of Web sites were visited at regular intervals for 12 months, and for each site
there is a record of the visits on which the site had changed from the prior visit.
 
The students spent the rest of the morning becoming familiar with the data, 
keeping in mind that the rate of change for a Web site was the main interest.
The data were provided in two different formats: one was a ragged array where for each Web site there was a set of times when a change was observed; 
and the other format was a data frame with a row corresponding to each detected 
change to a Web page, with a value for the time of detected change.
Advanced preparation with Grimes led us to the decision to provide these two distinct formats 
for the data so the students had more freedom in thinking about the problem in different ways 
without being constrained to follow a single path of analysis.

The students self-organized into groups of two and three and worked with
these representations of the data for the remainder of the morning. 
Before lunch they reported to everyone on their initial explorations.  
Several groups chose one of their plots to present and were given one minute
to describe an interesting feature of the data that was revealed in the figure.
The students had taken several different approaches and made many interesting observations. 
Grimes led a conversation with each group as they presented, and she assisted in 
uncovering various relevant features of the data. 
Students in the audience were invited to contribute their observations, 
confirm these findings, ask questions, or describe different results.
In this way, the students were guided to behave like a researcher, albeit in an accelerated 
fashion.
 
Students noticed many natural categories of Web site updating, e.g., many Web sites are 
created once and never change, while others are updated at regular intervals, and others 
change frequently but without any obvious pattern. 
With this new knowledge, the workshop reconvened after lunch and students continued their
investigations.  
This time they compared two groups of URLs, one that updated frequently and the other slowly. 
The goal was to investigate how similar they were with respect to the pattern of updates.  
This included a discussion led by Grimes on the exponential and Poisson distributions 
and their connection to the problem, i.e., changes to a site may follow a Poisson scatter.  
 
Throughout the day, in addition to exploring their ideas in small group conversations 
with Grimes, the 27 undergraduates were assisted by seven graduate students and three PhD statisticians. 
With such a low student-teacher ratio, students were able to quickly convert their 
ideas into working code and convey their discoveries to others.  
The next morning, the students recapped their work from the previous afternoon 
and presented their findings.
After this debrief, Grimes used the blackboard to uncover, with the students' insights from their
analysis, a problem with censoring that is inherent in the data.  
Web sites might be updated multiple times between visits but only one change is observed.  
Given the students' familiarity with the data, they understood the reasons for
adopting a more complex estimator of the rate of change based on the MLE~\citep{bib:Grimes} 
even though many did not have the related theoretical background. 

This iterative process with the researcher was highly choreographed, yet still open to creative ideas.
In this way, the students were figuring things out on their own, uncovering important issues and
discussing them with Grimes.  As the second day progressed,
students were increasingly independent and branched off in different directions. For example,
with the help of functions that made it easy to simulate the censored process, some students 
investigated and compared properties of the na{\"i}ve estimator that ignores the censoring 
and the improved MLE. 
Other students explored via a simulation study how to choose the optimal time between updates. 
When groups prepared and presented their findings, visualization was again the 
main vehicle to present their results.  Since they had not all been working on the same aspect 
of the problem, they were able to contribute different pieces to the story.

Grimes wrapped up the two days with a final presentation.  
She described a Bayesian method that she had
developed to decide how often to crawl each URL and gave the students a 
sense of what is possible with advanced study of statistics.  
At this time, she also spoke about her path from an
undergraduate major in anthropology and archaeology, 
where a semester abroad in Guatemala sparked an interest in quantitative methods for dealing with disparate data, 
to graduate school in statistics where she worked on nonlinear dimensionality reduction problems, 
to what it is like to work as a research statistician at Google.

Although the workshop is short, students are able to engage in the creative research process
and experience the excitement of making ``independent" discoveries using modern statistics
methods and the power of practical and computational training (albeit mostly guided by the researcher). 
By engaging students in the modern practice of statistical research, we hope to inspire them to seek more training and other research experiences at the undergraduate and graduate level.

\section{ADDITIONAL WORKSHOP ASPECTS}

\paragraph{Participants.}
The workshop brings together about 37 participants each year from around the country, 
typically this includes 25 undergraduate students, five graduate students, three researcher-presenters, 
two organizers, and two additional research statisticians.  
The undergraduate students come from a broad spectrum of
institutions and academic preparation. 
Across the 6 workshops, 146 undergraduates participated from 78 institutions and 27 states.
Additionally, more than half (77) were women.  
Once established, the program typically had 150 to 250 applications annually and,
of those admitted to the program, the acceptance rate was about 90\%.

In the admissions process, we looked for a balance of students in terms of computing 
background, statistics background, and institution. Some students were statistics majors
who had taken advanced courses in their major and others were majors in other fields who
had taken only one or two courses but saw statistics as an important asset to their future studies.
No computing skills were required, but we did ensure the group consisted of students 
with some experience with statistical software. 
 
Typically five graduate students participate in the workshop.   
They come from different universities and bring different perspectives and experiences 
about graduate school.   
In each of the two most recent offerings, one of the undergraduate participants from the
previous year was invited to return as a ``graduate student" assistant.
 
Each year the lead researchers include statisticians from academia and 
research labs in government and industry.
The latter bring a valuable non-academic perspective, both in the nature of the applications
they bring and also on career options. 
The main criteria for selection are that the researcher works closely with another discipline,
has tremendous enthusiasm for their work, excellent communication skills, and 
flexible teaching style. 

In addition to the organizers, we have routinely invited other researchers to join the workshop
for two or three days.  
These visitors have included Joe Blitzstein (Harvard University), Di Cook (Iowa State University), 
Nicholas Horton (Amherst College), David James (Bell Labs), and Deborah Swayne (Shannon Research Labs, AT\&T).  
Aside from assisting undergraduates, they also have given short lectures on statistical topics and
software demos. They also interact with the students via panels and informally, and 
they provide different perspectives on career paths and experiences.

\paragraph{Program length.}
The brevity of the program has many benefits. 
High-profile researchers are willing to volunteer for the program and dedicate
time to prepare for and participate in the workshop.
Additionally, the ESR exposes a large number of undergraduates (typically 25 each year) 
to statistics research, compared to typical REUs.  
Moreover, given the size and length of the program, we have been more willing to take risks in
admitting students, 
with the goal of having the biggest impact by including students who we think would gain a
lot from the program.  
Reciprocally,  for the students, the low commitment and opportunity cost for 
them to spend one week learning about
statistics research means that they are willing to take the risk of attending the workshop.
If they discover that the field is not right for them, then they have not dedicated their entire
summer to the program.  Many students report that they are able to attend the workshop 
in addition to participating in other summer programs, jobs, and courses.
Lastly, the participants receive roughly the same amount of information and
advice about graduate school that they would receive in a longer program.

Without a doubt, in a one-week program, students do not learn as much about specific statistical methods
or get the same extensive training in computing or visualization as they would in a 6- or 10-week
program. However, this is not our goal. We simply want participants to see the scope and
importance of statistics in a variety of contexts and to experience the challenge and 
excitement of addressing real-world questions with modern data analysis so they might 
be encouraged to take the next step in studying statistics.

\paragraph{Computing.}
The one-week program begins with a one-day, fast-paced introduction to the statistical programming
environment R~\citep{bib:R}. 
This introduction is both general and carefully crafted to prepare the students
for the needs of the following days. 
(See \texttt{www.stat.berkeley.edu/users/summer} for the reference 
materials supplied to the students.)
During this training we take the opportunity to explore interesting data sets and teach visualization. 
We have found that few students have received training in visualization, and this topic 
maintains the interest of those who are new to R and those who have extensive experience
with it. 
Additionally, we attempt to provide differentiated instruction so students can find 
practice problems that are appropriate to their level.
During the two-day projects themselves, 
the graduate students assist with the computing details so as it is not a
barrier to the students' creative expression, yet they are able to   
appreciate the power and need for computational skills.
And, experienced students typically learn new things, 
such as more sophisticated graphical functionality and computational approaches.

\paragraph{Information about graduate school.}
An important goal of ESR is to encourage students to consider graduate studies in statistical science
and to provide them with information about how to apply to graduate school, what graduate school is
like, and career opportunities in statistical science.  We organize three panel sessions on topics
related to graduate school and careers.  The first panel is an information session and group
discussion on the process of applying to graduate school.  Students receive general advice and 
materials on how to write a statement of purpose, who to ask for a letter of recommendation, funding
opportunities, how to get the most out of a site visit, etc.  More specific advice is also offered,
based on faculty experience on graduate admissions committees, about preparing for graduate school,
what graduate programs look for in an application, and also how to identify programs that are a good
fit for each student.  The second panel involves graduate students discussing their experience and
perspective about the difference between life as an undergraduate versus graduate student, the
process of selecting a graduate program and a PhD advisor, and student ``community."  
The final panel session includes statistical researchers working outside of a university setting, 
e.g.,  at industrial and national labs.  The panelists offer their views on these
non-university careers.  Each of these panels generates many questions from and engaged discussions
with the students and often provides eye-opening information to some about the possibilities of
graduate school.  Additionally, there are many informal opportunities over breaks and meals for
students to receive individual advice on preparing for, applying to, and selecting a graduate
school.

\paragraph{Variants.}
We have experimented with a few variations on the presentation of three two-day projects. 
For example, the first ESR included five topics, each for a single day.  
We found that one day did not give the students enough time to familiarize themselves with the problem and data.  
As a result, the students were mechanically solving the problem without having 
an opportunity to think of approaches
themselves and understand the implications of their discoveries and contribute ideas.  Additionally,
the context switching from one day to the next was mentally exhausting for them.  
We have had more success with two other variations. 

Most recently, we had only two topics for the week. Instead of a third topic, 
we included a project where students worked in groups on one of six data sets
that we provided.  
These were introduced on the first day of the workshop, and students had time to explore
them during the R tutorial.
They continued to work on the project during the second day, exploring interesting
features of the data. Then, the formal two-day sessions began on the third day.  
This schedule gave the students an opportunity to further hone their R skills in preparation for the research topics. 
On the last day, they completed their analysis and presented their findings. 
In evaluations, many students commented that they liked having their own separate project to
work on.  
Others noted that the continuity of working on their project throughout the week made it very
apparent how much their R skills had improved.     
	
In 2011 and 2013, we also offered a science version of the ESR
with Berkeley faculty leading the research topics. 
The format of these workshops was slightly different from the ESR.
Here there was a single theme for the week, such as the carbon cycle and sensor networks. 
These workshops included other activities, such as a poster session and having students 
design experiments and collect data. 
Like ESR, the common thread was working with data collected to address an
important, current research problem.

\section{REPORTED IMPACT}\label{sec:Impact}
Each year we have carried out end-of-program evaluations. 
We present here a summary of student feedback about the program that focuses on students' perceived benefits of the workshop. 
Overall, the students report that the material in ESR is very different from what they are exposed to in traditional coursework and they left the program with a much better understanding of the role of statistics in scientific discovery.
More specifically, students were asked what were the most valuable aspects of the workshop. 
Five main themes emerged from their responses. In decreasing order of mention these are: 
hands on experience with real data; exposure to modern statistics research; 
gaining expertise in R; 
access to faculty and graduate students; and information about graduate school.

When asked what were the least favorable aspects of the workshop, 30\% reported nothing was unfavorable.  
The rest listed issues that mainly fell into two areas.  
One related to the different levels of experience with R. 
Some students were frustrated with sitting through the introduction to the language
and others wished there was more time for preparation. 
The other problem raised was the level of technical detail that certain topics required.
For example,  we have found that the extensive background material needed for 
topics, such as genetics, can be a barrier to understanding the research problem.
We have found that it is important to get students working with the data quickly,
making discoveries, and offering insights on their own. 
This way they have a more rewarding experience despite the severe time limitation.

Students were also asked what surprised them about the workshop.  
Three themes emerged from their answers to this question.  
One was the importance of computing to modern statistical applications. 
As one student put it, he/she was surprised at 
``How important having good computing skills is for a statistician."  
And another student added a related note that he/she was surprised at 
``How many ways one can approach statistics problems visually."  
Another theme was the high quality of the speakers.
The students were very appreciative of the dedication of the experts who shared
their research problems with them.  
Also mentioned regularly by the students in response to this question was the group work and the supportive community 
created by the faculty and graduate students.  
The students enjoyed the collaborative, non-competitive environment. 

Finally, students were also asked: 
If recommending this workshop to a fellow student, 
what reasons would you give to him or her to participate? 
And, What reservations, if any, would you express to him or her about participating?
Below are representative reasons to participate: 
``The problems covered are a lot more interesting, mentally stimulating and applicable [than] 
what you see in classes."
``It will expand your knowledge of statistics and data analysis in a major way, both through the topics, and through collaboration with peers, grad students, and the amazing professors."
``Using real life data in engaging exercises."
``Insight on what sorts of cutting-edge research is being done in stats."
``To see how many applications statistics [has] and how it is not just a science but also an art."
``It is self-driven research unlike anything in a classroom"

As for the reservations they would express to someone, 
about one-third had no reservations and those that did made the following types of comments:
``Make sure you know a little R coming in because it can really make a difference."
``The workshop will be valuable only if you put in a lot of effort during the breakout sessions."
``Be prepared for a full week with not a ton of free time."

For a different perspective, 
we recently contacted the graduate student assistants from the past four offerings of the program.
We asked them to comment on the ways, if any,
the ESR has influenced their teaching and 
on what other ways they have benefited from the program.
Fourteen of  the 15 people contacted responded. 
They commented on the benefits of experiencing an alternative approach to teaching
that was more interactive and real-problem-oriented and on being exposed to 
researchers in other areas of statistics and seeing how they think about their research
problems. It was clear from their responses that they felt they had participated 
in a very different teaching and learning environment than previously experienced, 
and that they benefitted from exposure to this environment.
Below are representative comments from their evaluation: \\
``I think ESR definitely was a great experience for me to approach teaching from a more hands-on, open-ended perspective. Much of my previous teaching had been centered around set curriculum and going over pre-set problems, but what we did during the program helped me communicate with students and colleagues in a more creative and collaborative way - which encourages deeper thinking and discussion."
``I think one of the biggest impacts ESR had on my teaching is to recognize that while it is uncomfortable to me to give students open-ended problems, it is beneficial to their learning and it is exciting to see what directions they take the problems.  The program also taught me that it is good to sometimes give students messy data."
``The program gave me an opportunity to interact with students with many different statistical backgrounds and research interests. As I now collaborate frequently with social scientists, I am finding my previous experience in ESR quite helpful in my current work." 
``The program gave me experience to respond on the spot to all kinds of surprising students' questions. ...
[it] made me better in articulating my ideas/questions and understanding what other people were thinking when solving problems."
``When I first went on the job market for an academic job at a liberal arts institution, most of my teaching experience involved lecturing in a large-classroom setting.  Several hiring committees were intrigued by the ESR format and were happy I had experience facilitating an active learning environment."
``I learned new statistical concepts and application areas while I was TAing -- making me more well-rounded, and better equipped to make contributions to problems outside my research area."

In summary, the undergraduate and graduate students highly valued the experience of 
working with authentic data on current, relevant scientific problems in a collaborative
open-ended environment.  
We too believe these are essential elements of the program, and in the next section, 
we summarize the key features of the ESR that we think are responsible for creating this experience,
and we make recommendations for how to incorporate 
some of these features into the classroom so more students are exposed to authentic data analysis
processes.

\section{KEY FEATURES \& RECOMMENDATIONS}\label{sec:rec}

Our teaching and courses have benefited tremendously from
preparing, participating in, and experimenting with the ESR.
We have found our own  efforts to bring the ESR into our classroom have
helped create a higher level of student engagement, interest, and aptitude. 
From our experiences with the various versions of the workshop and from student evaluations,  
we have identified several aspects of ESR that we think are particularly important to its success,
and we provide a set of recommendations for ways to incorporate aspects of these key features
into ``regular" courses.   
These recommendations include both ideas for how individuals can change their 
courses and how we as a community of statistics faculty can bring about larger change.

\paragraph{1. The Research Problem.} 
For the ESR, we invite researchers  who are known 
for their active engagement in a scientific application,
their tremendous enthusiasm for their work, and 
exceptional communication skills.
The researchers' close connection to the application fosters 
enthusiasm for statistics among the students  as they see the 
relevance of the field in solving important problems at the frontiers of science. 
Moreover, the approach that the researchers take to engage students in 
the creative process of data analysis follows a non-traditional teaching practice  
that is more akin to  an investigatory process.  
In preparing for each ESR, we had the privilege to be in regular contact with the 
researchers in advance of the workshop.  
This preparatory work included reading the relevant papers describing the researchers' work,
exploring the data, and documenting our initial questions and thought 
process in this first exposure to the problem.
We acted as students during this preparatory  stage and our learning process
helped inform and shape the teaching and learning experience during the ESR. 

This experience aided us in developing an approach/philosophy for adapting
these projects into case studies and assignments for our courses.
These case studies are more focused on the scientific problem itself than
those typically found in, e.g., DASL~\citep{bib:DASL} which generally aim at
providing a brief example of a statistical method. 
Rather, they  are more in line with the context-laden open-ended case studies in~\cite{bib:NolanSpeed}.
However, they contain greater details on the statistical analysis 
(different possible approaches and statistical issues) and on computing and visualization. 
We have made some of these data and materials available for teaching
advanced undergraduate courses in~\cite{bib:NolanTempleLang} and its 
accompanying Web site \texttt{http://rdatasciencecases.org/}, and 
other materials for teaching introductory course are on the Web at \texttt{http://www.stat.berkeley.edu/users/summer/}.

While it can be difficult for instructors to find or access realistic, 
cutting-edge problems and very time consuming to work through the details,
especially without access to the researcher, we encourage instructors of 
statistics to cull problems from their own applied research or to 
collaborate with a local applied statistician or scientist to develop
a case study and make it available to the statistics community. 
The great advantage to developing a local application is that there is the possibility
of bringing in an expert, as with the ESR. 
For example, a local expert can be  invited to a class meeting to introduce the problem
and data and then invited back for a follow-up meeting to discuss student findings.

\paragraph{2. Visualization.}
We have found that structuring the initial stages of analysis around visualization creates 
a level playing field for the students and quickly engages them with the data. 
With visualization, students can uncover important aspects of a problem without 
needing knowledge of advanced methods. 
Despite their varied backgrounds, all students typically find that through a visualization
they can make a contribution that addresses the research problem.
From there, students head in different directions analyzing the data with more 
sophisticated statistical techniques depending on their preparation. 

Exploratory visualization is a vital element of all data analyses that
is rarely emphasized or explicitly taught in our courses. 
Often only a few simple types of visualizations are used in courses, such as
histograms, box plots, and scatter plots, and little or no attention is paid to the principles
of good graphics. Presentation graphics are important for making
convincing arguments, exploratory graphics are important for informing a data analysis,
and modern software tools have reduced the barrier to making rich, informative data visualizations.
For these reasons, we advocate that statistical graphics deserves a larger part in our curriculum. 
And importantly, students find it empowering and enjoyable to create informative and meaningful
visualizations.
See, e.g., \cite{bib:NolanPerrett} for examples of visualization assignments that
can be used in a spectrum of undergraduate  courses, and see
\texttt{http://datascience.ucdavis.edu/NSFWorkshops/Visualization/\\GraphicsPartI.pdf}
for an overview of material on graphics that we have included in our introductory and advanced
courses.

\paragraph{3. Computing.}  
The preparatory work mentioned in recommendation \#1 typically also included 
having us  build an R package that contains data, possibly in multiple formats, 
and supporting functions that make certain aspects of the analysis more 
convenient for the students.
In creating these packages, we developed a  sequence of activities 
that was highly choreographed, yet still permitted students to 
be creative in their analysis.

As noted in \cite{bib:ASAGuidelines} p.11, ``undergraduate statistics majors need facility with
computation to be able handle increasingly complex data and sophisticated approaches
to analyze it."
We would go so far as to assert that the statistics community must
treat computing as fundamental as basic mathematics and writing.
And, we would expect our students to have data manipulation skills.

Generally, many instructors are having positive experiences using R in introductory courses,
and in our experience, using R for scripting a data analysis can be easier for students 
than using a statistical calculator.
Also,  the growing popularity of RStudio~\citep{bib:RStudio} and 
approaches such as Project Mosaic~\citep{bib:mosaic} make using R even easier at 
the introductory level.   
Additionally, we have found that having students use dynamic 
documents with runnable code, e.g.,  knitr~\citep{bib:knitr} and R Markdown~\citep{bib:rmarkdown}
offers a workflow that helps organize code and text, which reduces the barrier to 
computational work in introductory courses. \cite{bib:BaumerEtAl} provide 
specifics and examples. A side benefit of this approach is that it 
can help instill a model of transparency and ethical practice of statistics.

\paragraph{4. Engagement.} 
The low student-teacher ratio and the freedom from assessment created an
open exchange of ideas between the undergraduates, graduates, and PhD statisticians.
Explicitly requiring the students to work in groups, asking them to have their own ideas 
about analyzing the data, expecting them all to contribute, 
and making it clear that there was no single correct answer, were some of the key features
that we believe helped foster their curiosity and gain confidence in expressing their ideas.
The undergraduates could always find someone to discuss an idea with
or to ask for assistance with programming.  Graduate students were able to take care of many
immediate issues and also help to identify more significant problems that required 
input from the researcher or organizers. This quick turn around created open, responsive
channels of communication that helped sustain the excitement of the data analysis.

We have found that we can partially create this atmosphere in our classes
through the use of technology and near-peer instructors.
Near-peers are students who have more advanced standing and have
previously taken the particular course.  
They act as instructional aides by assisting in lab sessions. 
Research shows that peer instruction increases student mastery of both conceptual reasoning 
and quantitative problem solving and increases student engagement~\citep{bib:CrouchMazur}. 
This approach can be particularly effective at large universities where low student-teacher 
ratios are not possible. 

We have also had some success using online forums for addressing questions about
projects and data analyses. We particularly like  Piazza (\texttt{https://piazza.com/}).
We organize our courses so that instructors, teaching assistants and near-peer instructors 
share in the responsibility of monitoring and responding to student posts, and
as the semester progresses, we reduce our responses with the expectation that
students fill in the gap and answer each others questions.  

There are many other possibilities for creating a community of student statisticians. 
For example, faculty can flip the classroom, where there is more time for student-student
and faculty-student conversation on how to approach a data analysis problem 
in the classroom because students are receiving the more traditionally delivered 
material outside the classroom.  
As another option, faculty can sponsor a DataFest~\citep{bib:DataFest} at their institution where 
students work intensively in groups for three days on a real-world project.
Possibly, student clubs can succeed here as well.

\paragraph{5. Advanced methods.}  
In the ESR, after the students have worked with the data and have an understanding
of the research question, the researcher introduces an advanced method to analyze the data, 
such as spline smoothing, recursive partitioning, and empirical Bayes. 
This introduction is in the context of solving the current problem and from an intuitive point of view,
rather than a more abstract, rigorous mathematical approach. 
In this context, students are excited about seeing how modern methods can
be used to solve important real world problems.  
The students are given a basic understanding of 
how the method works and why it is useful in the particular setting, but they are also
well aware that further study of statistics is essential to understanding how best to
employ these tools.

We advocate that our undergraduate curriculum needs to introduce modern, advanced (and fun!) statistical methods into introductory courses. 
Typically, our courses focus on topics  such as histograms,
$t$-tests, and simple linear models, but why not also include one or more modern 
topics that are easy to understand at an intuitive and/or algorithmic level and that
can excite students about statistics and attract them to the field?
These methods  can be incorporated into case studies that use more basic methods and
so bring the teaching of statistics closer to the practice of statistics. 
A small change such as this has the potential to make a large impact on student 
interest in and perception of our field.
Moreover, if the concepts behind testing and inference are embedded in this larger framework, 
we believe that students will better understand and properly use statistics.

\section{CONCLUSION}
In this article, we have described a program for undergraduates that aims to create a
rich and vibrant experience working with modern, authentic, statistics research problems.  
We have attempted to convey the unique aspects of the program with the hope
that it will spark ideas and lead to change in our undergraduate statistics introductory
courses and major programs. 
The first three guiding principles of the 2014 ASA Guidelines for undergraduate programs in 
statistical science are: the scientific method and its relation to the statistical problem solving
cycle; real applications; and focus on problem solving.  
The  ESR provides insights into how we might improve our curricular activities to
follow these guiding principles. 
For example, we can give students early practice with the interplay between
questions, answers and statistics and with authentic data analysis.
We also can update curricular topics to increase emphasis on data visualization and  
incorporate modern methods into introductory classes. 
Furthermore, there are opportunities with near-peer instruction, online discussion boards,
etc. to foster a community of engaged student learners. 

Finally, faculty development appears at the top of the list of  ``next steps'' in the ASA Guidelines,
which calls for creating and sharing materials, such as those mentioned in Section~\ref{sec:rec}.
We further advocate creating opportunities for faculty to participate in inquiry-based 
approaches to teaching and approaches for bringing statistical problem solving 
into the undergraduate classroom, similar to the graduate students' experience in the ESR.
One possibility would be to develop an ESR-like experience  for faculty 
where they have the opportunity to create materials to use in their classrooms and 
share with others. 
If statistics is to remain a vital field, then we must modernize our teaching, both 
the topics we cover and our approach to teaching them. 
Statistics educators are a key piece of this change.

\singlespacing
\bibliography{ESR}

\begin{thebibliography}{15}
\providecommand{\natexlab}[1]{#1}
\providecommand{\url}[1]{\texttt{#1}}
\expandafter\ifx\csname urlstyle\endcsname\relax
  \providecommand{\doi}[1]{doi: #1}\else
  \providecommand{\doi}{doi: \begingroup \urlstyle{rm}\Url}\fi

\bibitem[Allaire(2015)]{bib:rmarkdown}
J.~J. Allaire.
\newblock {\textit{rmarkdown}: Dynamic Documents for R}.
\newblock \texttt{http://cran.r-project.org/web/packages/rmarkdown}, 2015.
\newblock R~package version~0.5.1.

\bibitem[{ASA}(2014)]{bib:ASAGuidelines}
{ASA}.
\newblock \emph{2014 Curriculum Guidelines for Undergraduate Programs in
  Statistical Science}.
\newblock Alexandria, VA, 2014.
\newblock \texttt{http://www.amstat.org/education/\\curriculumguidelines.cfm}.

\bibitem[Baumer et~al.(2014)Baumer, Cetinkaya-Rundel, Bray, Loi, and
  Horton]{bib:BaumerEtAl}
B.~Baumer, M.~Cetinkaya-Rundel, A.~Bray, L.~Loi, and N.~J. Horton.
\newblock R markdown: Integrating a reproducible analysis tool into
  introductory statistics, 2014.
\newblock \texttt{http://arxiv.org/pdf/1402.1894.pdf}.

\bibitem[Cetinkaya-Rundel and Stangl(2013)]{bib:DataFest}
M.~Cetinkaya-Rundel and D.~Stangl.
\newblock {A Celebration of Data}.
\newblock \emph{{The American Statistician}}, 23\penalty0 (3):\penalty0 43--46,
  2013.

\bibitem[Crouch and Mazur(2001)]{bib:CrouchMazur}
C.~Crouch and E.~Mazur.
\newblock {Peer Instruction: Ten years of experience and results}.
\newblock \emph{{Am. J. Phys.}}, 69:\penalty0 970--977, 2001.

\bibitem[{DASL Project}(2014)]{bib:DASL}
{DASL Project}.
\newblock {Data and Story Library (DASL)}.
\newblock \texttt{http://lib.stat.cmu.edu/DASL/\\DataArchive.html}, 2014.

\bibitem[Grimes et~al.(2008)Grimes, Ford, and Tassone]{bib:Grimes}
C.~Grimes, D.~Ford, and E.~Tassone.
\newblock {Keeping a Search Engine Fresh: Risk and optimality in estimating
  refresh rates for web pages}.
\newblock In \emph{{Proceedings of the INTERFACE}}. 2008.
\newblock \texttt{http://www.niss.org/interface2008-talks}.

\bibitem[Nolan and Perrett(2014)]{bib:NolanPerrett}
D.~Nolan and J.~Perrett.
\newblock Copying the masters and other techniques for learning data
  visualization, 2014.
\newblock arXiv:1503.00781.

\bibitem[Nolan and Speed(2001)]{bib:NolanSpeed}
D.~Nolan and T.~P. Speed.
\newblock \emph{{Stat Labs: Mathematical Statistics through Applications}}.
\newblock {Springer}, 2001.

\bibitem[Nolan and {Temple Lang}(2015)]{bib:NolanTempleLang}
D.~Nolan and D.~{Temple Lang}.
\newblock \emph{{Data Science in R: A Case Studies Approach to Computational
  Reasoning and Problem Solving Mathematical Statistics through Applications}}.
\newblock {CRC Press}, 2015.

\bibitem[Pruim et~al.(2014)Pruim, Kaplan, and Horton]{bib:mosaic}
R.~Pruim, D.~Kaplan, and N.~J. Horton.
\newblock {\textit{mosaic}: Project MOSAIC statistics and mathematics teaching
  utilities}.
\newblock \texttt{http://cran.r-project.org/web/packages/mosaic/}, 2014.
\newblock R~package version~0.9.1-3.

\bibitem[{R Development Core Team}(2012)]{bib:R}
{R Development Core Team}.
\newblock \emph{R: A Language and Environment for Statistical Computing}.
\newblock Vienna, Austria, 2012.
\newblock \texttt{http://www.r-project.org}.

\bibitem[{RStudio}(2013)]{bib:RStudio}
{RStudio}.
\newblock \emph{RStudio: Integrated development environment for R; Version~
  0.98.978}.
\newblock Boston, MA, 2013.
\newblock \texttt{ https://www.rstudio.com}.

\bibitem[Speed(1986)]{bib:Speed}
T.~P. Speed.
\newblock {Questions, Answers, and Statistics}.
\newblock In \emph{Proceedings of the International Conference on Teaching
  Statistics 2}. 1986.

\bibitem[Xie(2014)]{bib:knitr}
Y.~Xie.
\newblock {\textit{knitr}: A General-Purpose Package for Dynamic Report
  Generation in R}.
\newblock \texttt{http://cran.r-project.org/web/packages/knitr}, 2014.
\newblock R~package version~1.8.

\end{thebibliography}

\end{document}